     \documentstyle[12pt]{article}

\begin{document}

\begin{flushright}
CERN-TH. 228-95
\end{flushright}

     {\center{{\bf{\Large QCD Pressure at Two Loops in the Temporal
     Gauge}}}
     \\
     \center{George Leibbrandt\footnotemark\ \\
     Theoretical Physics Division, CERN, CH-1211 Geneva 23 \\
     Switzerland}
     \\
     \center{Mark Staley \\ Department of Physics, University of
Guelph,
     and
     Guelph-Waterloo Program for Graduate Work in Physics, \\
Guelph
     Campus, Guelph,
     Ontario N1G 2W1}

     \addtocounter{footnote}{0}\footnotetext{Permanent address:
Department
     of Mathematics and Statistics, University of Guelph, \\
Guelph,
     Ontario, N1G 2W1; E-mail: gleibbra@msnet.mathstat.uoguelph.ca}

     \vfill

     \begin{abstract}
     \noindent
     We apply the method of \underline{zeta} functions, together
with the
     $n_\mu^*$-prescription for the temporal gauge, to evaluate
the
     thermodynamic
     pressure in QCD at finite temperature $T$. Working in the
     imaginary-time
     formalism and employing a special version of the unified-gauge
     prescription,
     we show that the pure-gauge contribution to the pressure at
two loops
     is given
     by $P_2^{\mbox{{\scriptsize gauge}}} = -(g^2/144)N_cN_gT^4$,
where
     $N_c$ and $N_g$
     denote the number of colours and gluons, respectively.  This
result
     agrees
     with the value in the Feynman gauge.
     \end{abstract}
     }
     \vfill
     \vfill\eject

     \baselineskip=20pt

     \section{Introduction}

     The temporal gauge is a physical, i.e. ghost-free gauge which
belongs
     to the class of axial-type gauges characterized by the
constraint
     \begin{equation}
     n^\mu A_\mu^a(x) = 0, \ \ \mu=0,1,2,3,\ldots,
     \end{equation}
     where $A_\mu^a(x)$ is a massless Yang-Mills field, and
$a=1,2,\ldots,
     N^2-1$,
     for SU(N).  These axial-type gauges also include the pure
axial gauge
     ($n^2<0$), the planar gauge ($n^2<0$), and the light-cone
gauge
     ($n^2=0$).
     In the temporal gauge, the fixed vector $n_\mu$ is time-like:
     $n^2 = n_0^{\,2} - {\bf n}^2 > 0$.  Today, the temporal gauge
is as
     useful as
     it was 65 years ago in the quantization of the Maxwell-Dirac
field by
     Weyl in
     1929 \cite{Weyl} and Heisenberg and Pauli in 1930
\cite{Heisenberg}.
     Only the degree of complexity has increased: today's temporal
gauge is
     applied
     in such sophisticated areas as the vacuum tunneling by
instantons
     \cite{Rossi},
     Nicolai maps \cite{Claudson,Bern}, one-loop thermodynamic
potentials
     \cite{Actor},
     and especially in quark-gluon plasma studies \cite{Kajantie}
at {\em
     finite}
     temperature $T$.  Of course, there are good reasons for this
     popularity.

     Due to the space-time asymmetry at finite temperature in the
     imaginary-time
     formalism, the thermodynamic equations are no longer symmetric
in
     4-space,
     but are only invariant under spatial rotations.  Accordingly,
since
     the temporal
     gauge-fixing condition $n_0A_0^a(x) = 0$ leaves the spatial
components
     of
     $A_{\mu}^a(x)$ unconstrained,
     the temporal gauge may indeed be regarded as the ``appropriate
gauge''
     for this
     particular boundary-value problem.

     During the last ten years, finite-temperature calculations in
the
     temporal gauge have been performed both in the real-time
formalism and
     in the
     imaginary-time formalism.  Kapusta and Kajantie
\cite{Kajantie}, for
     instance,
     employed the imaginary-time formalism to evaluate the leading
terms in
     the
     $1/T$-expansion for the two-point function (static limit).
The
     finite-temperature case was also examined by Landshoff and
James
     \cite{James},
     as well as by Brandt, Frenkel and Taylor \cite{Brandt}.
Working in
     the Feynman
     gauge, Brandt, Frenkel and Taylor utilized the method of
     $\zeta$-functions
     to derive the complete $1/T$-expansion for the one-loop
Yang-Mills
     self-energy.

     In 1994, the present authors developed a general procedure
\cite{L-S}
     for doing
     perturbative calculations in the temporal gauge at both zero
($T=0$)
     and
     finite temperature ($T \not= 0$). The procedure hinged on a
special
     version of
     the $n_\mu^*$-prescription, originally designed for the
spurious
     poles of the
     temporal-gauge propagator (cf. Eq.~(\ref{e:propagator})).  No
problems
     were
     encountered for \underbar{zero} temperature, but for $T \not
= 0$ a
     different
     approach was needed.  In order to simplify the computation of
     Matsubara
     frequency sums, it became necessary to replace the traditional
contour
     method
     by the method of zeta functions.  The latter method is
generally known
     to facilitate the computation of high-temperature expansions
to any
     order in
     $1/T$.

     Application of the $\zeta$-function technique, together with
the
     $n_\mu^*$-prescription for the temporal gauge, enabled us to
     evaluate the
     complete $1/T$-expansion for the self-energy
     $\Pi_{00}^{ab}(k_4=0, {\bf k})$\cite{L-S}, and to derive the
Debye
     chromo-electric screening length $m_{\mbox{el}}$ in the
infrared limit,
     $m_{\mbox{el}}^2 \equiv \Pi_{00}^{ab}(0, {\bf k})
     \vline_{{\bf k}^2 = m_{\mbox{{\tiny el}}}^2}$\cite{Rebhan}.

     The purpose of the present article is to apply this successful
     procedure
     \cite{L-S} to another important quantity, namely the
thermodynamic
     pressure.  Working in the imaginary-time formalism, we shall
evaluate
     the
     thermodynamic pressure at finite temperature to one and two
loops.

     \section{Basic Tools}

     The temporal-gauge propagator for Yang-Mills theory at finite
     temperature
     reads as follows \cite{L1,L2}:
     \begin{equation}
     G_{\mu\nu}^{ab}(p) =
     \frac{-i\delta^{ab}}{(2\pi)^{2\omega}(p^2+i\epsilon)}
     \Big [ g_{\mu\nu} - \frac{p_{\mu}n_{\nu} +
p_{\nu}n_{\mu}}{p\cdot n}
     +\frac{n^2p_{\mu}p_{\nu}}{(p\cdot n)^2} \Big ], \ \ n^2>0, \
\epsilon
     > 0,
     \label{e:propagator}
     \end{equation}
     where $2\omega$ denotes the dimensionality of complex
space-time; in
     four
     dimensions, $g_{\mu\nu}$ = diag(+1, -1, -1, -1), $\mu, \nu =
0,1,2,3$.
     For
     $n_{\mu} = (1,0,0,0)$, the propagator (\ref{e:propagator})
reduces to
     \begin{eqnarray}
     && G_{ij}^{ab}(p) =
     \frac{-i\delta^{ab}}{(2\pi)^{2\omega}(p^2+i\epsilon)}
     \Big [ -\delta_{ij} + \frac{p_ip_j}{(p_0)^{2}} \Big ],
     \ \ i,j=1,2,3, \nonumber \\
     && G_{00}^{ab}(p) = G_{0i}^{ab}(p) = G_{i0}^{ab}(p) = 0.
     \label{e:propsimp}
     \end{eqnarray}

     The spurious double pole at $p_0=0$ in Eq.~(\ref{e:propsimp})
may be
     treated
     by a special version of the unified-gauge prescription for
axial-type
     gauges
     \cite{L2,L3}, namely
     \begin{eqnarray}
     \frac{1}{p_0} \vline^{\mbox{temp}} &&= \lim_{\epsilon \to 0}
     \frac{p_0}{p_0^2+i\epsilon}, \ \ \epsilon > 0; \nonumber \\
     \frac{1}{(p_0)^2} \vline^{\mbox{temp}} &&= \lim_{\epsilon \to
0}
     \Big (\frac{p_0}{p_0^2+i\epsilon}\Big )^2, \ \ \epsilon > 0.
     \label{e:prescription}
     \end{eqnarray}
     Note that prescription (\ref{e:prescription}) is
\underbar{causal}
     and well defined for all $p_0 = 2\pi inT$, $n=0, 1,2, \ldots$,
$T$
     being the
     temperature.  In particular, no ambiguities arise for the case
$n=0$.

     In addition to the formulas (\ref{e:propsimp}) and
     (\ref{e:prescription}),
     loop calculations at finite temperature also require the
replacement
     \begin{equation}
     \int \frac{d^4p}{(2\pi)^4} \Longrightarrow \frac{i}{\beta}
     \sum_{n=-\infty}^{n=+\infty} \int \frac{d^3{\bf p}}{(2\pi)^3}.
     \end{equation}
     Here $\beta \equiv 1/(kT)$, and $p_0 \equiv i\omega_n$, with
$\omega_n
     = 2\pi nT$;
     it is customary to equate Boltzmann's constant $k$ to unity.
Finally,
     we note
     that the \underbar{components} of $n_{\mu} = (n_0, n_1, n_2,
n_3)$
     have the following
     structure in the temporal gauge \cite{L3}:
     \begin{eqnarray}
     &&n_{\mu} = (n_0, {\bf n}_{\perp}, -i|{\bf n}_{\perp}|),
     \ \ n_0 \not = 0, \ {\bf n}_{\perp} = (n_1, n_2); \nonumber \\
     &&n_{\mu}^* = (n_0, {\bf n}_{\perp}, +i|{\bf n}_{\perp}|),
     \label{e:n}
     \end{eqnarray}
     so that
     \begin{equation}
     \lim_{|{\bf n}_{\perp}| \to 0} n_{\mu} = (1,0,0,0), \ \ n_0
\equiv
     1.
     \end{equation}

     \section{The Q.C.D. Pressure to One Loop}

     Since the pressure $P$ is defined in terms of the partition
function
     $Z$,
     \begin{equation}
     P = T\frac{\partial \ln Z}{\partial V}, \ \ T \equiv
     \mbox{temperature},\ \ V \equiv \mbox{volume},
     \end{equation}
     we first need to calculate $Z$ for pure gauge theory
\cite{Kapusta}:
     \begin{equation}
     Z_{\hbox{gauge}} = \int \big [dA_{\mu} \big ] \det(n\cdot
\partial)
     e^S;
     \end{equation}
     $[\ ]$ denotes a product over all field configurations, with
the
     action
     $S$ given by
     \begin{eqnarray}
     S &=& \int_0^{\beta}d\tau \int d^3x \Big [ -\frac{1}{4}
F^{\mu\nu}
     F_{\mu\nu} - \frac{(n\cdot A)^2}{2\alpha} \Big ], \nonumber \\
     &=& \frac{1}{2} \int_0^{\beta} d\tau \int d^3x A^{\mu}
     \Big ( \partial^2\delta_{\mu\nu} -
\partial_{\mu}\partial_{\nu}
     - \frac{n_{\mu}n_{\nu}}{\alpha\beta^2} \Big ) A^{\nu}.
     \label{e:S}
     \end{eqnarray}
     We shall keep $n_\mu$ general for now (cf. Eqs.~(\ref{e:n})),
taking
     the
     limit $|{\bf n_{\bot}}| \to 0$ at
a
later
time.
Notice
that periodicity
     of the
     boundary condition causes the boundary term in Eq. (\ref{e:S})
to
     disappear.
     Introducing the Grassmann
     variables $C$ and $\bar C$ ($\bar C$ is the
     hermitian conjugate of $C$), we may write the determinant
     $\det (n \cdot \partial )$ as follows:
     \begin{equation}
     \det (n\cdot \partial) = \int \big [ d \bar C \big ] \big [ dC
\big ]
     \exp \Big ( \int_0^{\beta} d\tau \int d^3x \bar C
\partial\!\cdot\! n
     C \Big ).
     \end{equation}
     It is convenient now to apply a discrete Fourier
transformation in
     Euclidean
     space (the continuous limit will be taken later in the
calculation):
     \begin{equation}
     C({\bf x}, \tau) = \frac{1}{V^{\frac{1}{2}}} \sum_n \sum_{{\bf
p}}
     e^{i({\bf p}\cdot {\bf x} + \omega_n\tau)} \tilde C_n({\bf
p}), \ \
     \omega_n = (2n+1)\pi T,
     \end{equation}
     where $\tilde C_n({\bf p})$ is the Fourier transform of
$C({\bf x},
     \tau)$,
     ${\bf p}$ and $n$ label the momentum and energy quantum
numbers,
     respectively,
     and where the volume
     factor $V$ is necessary for proper normalization of the
     transformation. The frequency $\omega_n$ is odd here, because
$C$ is a
     fermion field.
     For the gauge field $A^{\mu}({\bf x}, \tau)$, we have:
     \begin{equation}
     A^{\mu}({\bf x}, \tau) = \sqrt{\frac{\beta}{V}}
     \sum_n \sum_{{\bf p}} e^{ i({\bf p}\cdot {\bf x} +
\omega_n\tau) }
     A_n^{\mu}({\bf p}), \ \ \omega_n = 2n\pi T.
     \end{equation}
     The extra factor of $\sqrt{\beta}$ is needed to keep the
     action dimensionless. Thus
     \begin{eqnarray}
     Z_{\mbox{{\small gauge}}} &=& \int \big [ dA^{\mu} \big ] \big
[ d
     \bar C \big ] \big [ dC \big ]
     \exp \Big \lbrace \int_0^{\beta} d \tau \int d^3 x \bar C
n\!\cdot\!
     \partial
     C \Big \rbrace \nonumber \\
     && \ \ \exp \Big \lbrace \int_0^{\beta} d\tau \int d^3 x
     \frac{\beta}{2V}
     \sum_{n,{\bf p}}\sum_{n^{\prime},{\bf p^{\prime}}}
     e^{ i(\omega_n+\omega_{n^{\prime}})\tau} e^{i({\bf p} + {\bf
     p}^{\prime})
     \cdot {\bf x}} A_n^{\mu}({\bf p})A_{n^{\prime}}^{\nu}({\bf
     p}^{\prime})
     \nonumber \\ && \ \ \ \ \ \ \Big ( -p^2 \delta_{\mu\nu} +
     p_{\mu}p_{\nu}
     - \frac{n_{\mu}n_{\nu}}{\alpha\beta^2} \Big ) \Big \rbrace, \\
     & & \nonumber \\
     &=& \int \big [ dA^{\mu} \big ] \big [ d\bar C \big ] \big [
dC \big ]
     \exp \Bigg \lbrace \int_0^{\beta} d\tau \int d^3 x \frac{1}{V}
     \sum_{n,{\bf p}}\sum_{n^{\prime},{\bf p^{\prime}}}
     e^{-i({\bf p} \cdot {\bf x} + \omega_n \tau)} \nonumber \\
     && \ \ \ \ \tilde{ \bar C}_n({\bf p}) (i{\bf p}^{\prime})
     e^{i({\bf p}^{\prime} \cdot {\bf x} + \omega_n^{\prime} \tau)}
     \tilde C_{n^{\prime}}({\bf p}^{\prime}) \Bigg \rbrace
     \exp \Bigg \lbrace \frac{\beta}{2V} (\beta V) \sum_{n,{\bf p}}
     A_n^{\mu}({\bf p}) A_{-n}^{\nu}(-{\bf p}) \nonumber \\
     && \ \ \ \ \Big ( -p^2 \delta_{\mu\nu}
     + p_{\mu}p_{\nu} - \frac{n_{\mu}n_{\nu}}{\alpha\beta^2} \Big
) \Bigg
     \rbrace, \\
     & & \nonumber \\
     &=&  \int \big [ d\bar C \big ] \big [ dC \big ] \exp \Bigg
\lbrace
     \frac{\beta V}{V}\sum_{n,{\bf p}} \tilde{ \bar C}_n({\bf p})
(i{\bf
     p})
     e^{i({\bf p} \cdot {\bf x} + \omega_n \tau)}
     \tilde C_{n}({\bf p}) \Bigg \rbrace \nonumber \\
     &\ & \int \big [ dA^{\mu} \big ] \exp \Bigg \lbrace
\frac{1}{2}
     \beta^2
     \sum_{n,{\bf p}} A_n^{\mu}({\bf p}) {A_n^{\nu}}^*({\bf p})
\nonumber
     \\
     && \ \ \ \ \Big ( -p^2 \delta_{\mu\nu}
     + p_{\mu}p_{\nu} - \frac{n_{\mu}n_{\nu}}{\alpha\beta^2} \Big
)\! \Bigg
     \rbrace,
     \end{eqnarray}
     where we have used $A_{-n}^{\mu}(-{\bf p}) =
{A_n^{\mu}}^*({\bf p})$,
     since
     $A^{\mu} ( {\bf x}, \tau)$ is real.  After further
simplification, we
     get:
     \begin{eqnarray}
     Z &=& \int \big [ d\bar C \big ] \big [ dC \big ]
     \exp \big \lbrace (\bar C, FC) \big \rbrace \int \big [
dA^{\mu} \big
     ]
     \exp \big \lbrace -\frac{1}{2} (A^{\mu} D_{\mu\nu} A^{\nu})
\big
     \rbrace,
     \nonumber \\
     &=& \det (F) \big ( \det (D_{\mu\nu}) \big )^{-\frac{1}{2}};
     \end{eqnarray}
     here
     \begin{eqnarray}
     && F = i\beta p\!\cdot\! n, \nonumber \\
     && D_{\mu\nu} = \big ( p^2\delta_{\mu\nu} - p_{\mu}p_{\nu}
     + \frac{n_{\mu}n_{\nu}}{\alpha \beta^2} \big ) \beta^2.
     \end{eqnarray}
     The determinants are calculated by summing over space-time
indices as
     well as
     field indices.  Further manipulation leads to
     \begin{eqnarray}
     \ln Z &=& \ln \det (\beta p\!\cdot\! n)-\frac{1}{2} \ln\det
     (D_{\mu\nu}), \nonumber \\
     &=& \ln\det (\beta \omega_n ) - \frac{1}{2} \ln \det
(D_{\mu\nu}),
     \end{eqnarray}
     with
     \begin{equation}
     D_{\mu\nu} = \left [ \matrix{
     p^2 - p_0^2 + \frac{1}{\alpha\beta^2} & -p_0p_1 & -p_0p_2 &
-p_0p_3
     \cr
     -p_0p_1 & p^2-p_1^2 & -p_1p_2 & -p_1p_3 \cr
     -p_0p_2 & -p_1p_2 & p^2-p_2^2 & -p_2p_3 \cr
     -p_0p_3 & -p_1p_3 & -p_2p_3 & p^2-p_3^2 \cr} \right ] \beta^2.
     \end{equation}
     In the temporal gauge,
     \begin{equation}
     \det \big ( D_{\mu\nu} \big ) = \frac{1}{\alpha}
\beta^6p^4\omega_n^2.
     \end{equation}
     Hence
     \begin{eqnarray}
     \ln Z &=&  \ln \det \big ( \beta \omega_n\big ) - \frac{1}{2}
\ln \det
     \big (
     \frac{1}{\alpha} \beta^6 p^4 \omega_n^2 \big ), \nonumber \\
     &=& T\!r \ln \big ( \beta^2 \omega_n^2 \big )
     - T\!r \ln \big ( \frac{1}{\alpha} \beta^6 p^4 \omega_n^2 \big
),
     \nonumber \\
     &=& \Big ( \prod_n \prod_{{\bf p}} \big [ \beta^2 (\omega_n^2
+ {\bf
     p}^2) \big ]^{-1} \Big )
     + \sum_{n,{\bf p}} \ln \sqrt{\alpha}.
     \label{e:bla}
     \end{eqnarray}
     Absorbing the second term in Eq.~(\ref{e:bla}) into the
overall
     normalization
     of $Z$, we are left with the expression
     \begin{equation}
     \ln Z = -\sum_n \sum_{{\bf p}} \ln \big [ \beta^2 (\omega_n^2
+ {\bf
     p}^2 ) \big ],
     \end{equation}
     which may be further reduced by applying the formulas
\cite{Big1}:
     \begin{eqnarray}
     \ln \big [ (2\pi n)^2 + \beta^2 \omega^2 \big ] &=&
     \int_1^{\beta^2\omega^2}
     \frac{d\theta^2}{\theta^2+(2\pi n)^2} + \ln \big [ 1+(2\pi
n)^2 \big
     ], \nonumber \\
     \sum_{n=-\infty}^{+\infty} \frac{1}{n^2 +
(\frac{\theta}{2\pi})^2}
     &=& \frac{2\pi^2}{\theta}
     \Big ( 1+\frac{2}{e^{\theta} - 1} \Big ).
     \end{eqnarray}
     Finally, incorporating the $\beta$-independent term into the
     normalization of
     $Z$, and replacing $\stackrel{\sum}{{\bf p}}$ by
     \begin{equation}
     \sum_{{\bf p}} \longrightarrow \frac{V}{(2\pi)^3} \int d^3
{\bf p},
     \end{equation}
     we find that
     \begin{equation}
     \ln Z = 2V \int \frac{d^3{\bf p}}{(2\pi )^3} \big [
-\frac{1}{2} \beta
     \omega
     - \ln (1-e^{-\beta \omega} ) \big ], \ \ \omega = |{\bf p}|.
     \label{e:catsmeow}
     \end{equation}
     Eq.~(\ref{e:catsmeow}), which is just the usual black-body
radiation
     formula for the \underbar{two} physical
     gluon polarizations, agrees with a similar calculation in the
     Feynman gauge \cite{Kapusta}.  We now proceed with the more
     challenging
     task of computing the thermodynamic pressure at two loops.

     \section{The Q.C.D. Pressure at Two Loops}

     Higher-order terms in the partition function may be evaluated
by
     forming bubble diagrams and applying
     the usual Feynman rules.
     The pure gauge contribution to the pressure at two loops is
contained
     in the
     diagrams of Figure 1.  The first of these is called the oyster
     diagram, the
     second the bowtie diagram.
     The thermodynamic pressure was the first
     gauge-invariant, physical quantity to be calculated in
     finite-temperature
     Q.C.D. at the two loop level. The computation was carried out
by
     Kapusta \cite{Kapusta79} in the Feynman gauge who obtained the
     following
     result:
     \begin{equation}
     P_2^{\mbox{{\small glue}}} = -\frac{g^2}{144} N_cN_g T^4,
     \label{e:P2Feynman}
     \end{equation}
     where $N_c$ is the number of colours, and $N_g = N_c^{\ 2} -1$
the
     number
     of gluons.  $N_c$ and $N_g$ arise from counting the number of
     particles
     participating in the interaction.  As stated in the
     \underbar{Introduction},
     our aim is to calculate the 2-loop pressure in the
noncovariant
     temporal gauge,
     by using the method of $\zeta$-functions, along with a special
version
     of the
     unified-gauge prescription.

     Applying the Feynman rules of Section 2 to the oyster diagram
in
     Figure 1,
     we obtain:
     \begin{eqnarray}
     P_2^{\mbox{{\small oyster}}} &=& g^2N_cN_g
T\sum_{n=-\infty}^{+\infty}
     \int\frac{d^3{\bf p}}{(2\pi)^3} T\sum_{l=-\infty}^{+\infty}
     \int\frac{d^3{\bf k}}{(2\pi)^3} \frac{1}{(p+k)^2}
\frac{1}{p^2}
     \frac{1}{k^2} \nonumber \\
     &\ & \cdot \Big [ -\delta_{\mu\sigma}(p+2k)_{\tau} +
     \delta_{\sigma\tau}
     (2p+k)_{\mu}-\delta_{\tau\mu}(p-k)_{\sigma} \Big ] \nonumber
\\
     &\ & \cdot \Big [
-\delta_{\sigma^{\prime}\nu}(p+2k)_{\tau^{\prime}}
     -\delta_{\nu\tau^{\prime}}(p-k)_{\sigma^{\prime}}
     + \delta_{\tau^{\prime}\sigma^{\prime}}(2p+k)_{\nu} \Big ]
\nonumber
     \\
     &\ & \cdot \Bigg [ \delta^{\sigma\sigma^{\prime}} +
     \frac{(p+k)^{\sigma}(p+k)^{\sigma^{\prime}}}{(p_4+k_4)^2}
\Bigg ]
     \nonumber \\
     &\ & \cdot \Bigg [ \delta^{\tau\tau^{\prime}}
     + \frac{p^{\tau}p^{\tau^{\prime}}}{p_4^{\ 2}} \Bigg ]
\nonumber \\
     &\ & \cdot \Bigg [ \delta^{\mu\nu} +
\frac{k^{\mu}k^{\nu}}{k_4^{\ 2}}
     \Bigg ],
     \end{eqnarray}
     where $p_4 = 2\pi nT$, $k_4 = 2\pi lT$, and $\mu, \nu, \tau,
     \tau^{\prime},
     \sigma, \sigma^{\prime}$ are only summed over 1,2,3,
     while $n=1,2, \ldots$ and $l=1,2,\dots$.  Multiplication gives
rise to
     a total
     of 720 terms.  Contracting indices we find that the resulting
     expression
     reduces to 63 tadpole-like integrals, many of which are equal
to zero.
      The
     contribution from the oyster diagram is, therefore, given by
     \begin{equation}
     P_2^{\mbox{{\small oyster}}} = g^2N_cN_g
T\sum_{n=-\infty}^{+\infty}
     T\sum_{l=-\infty}^{+\infty} \sum_{i=1}^{63} I_i.
     \label{e:Bigsum}
     \end{equation}
     The integrals $\lbrace I_i \rbrace$ are summarized in the
Appendix.
     They are related to the familiar tadpole integrals which arise
in
     massless
     theories such as Q.C.D.  Let us briefly discuss this important
class
     of
     integrals.

     A typical example, at zero temperature, is the tadpole
integral $I$,
     \begin{equation}
     I = \int\frac{d^4 p}{(2\pi)^4} \
\frac{1}{p^2}.
     \label{e:tadpole}
     \end{equation}
     In the context of dimensional regularization, such integrals
may be
     formally
     set to zero \cite{C-L}. The \underbar{finite-temperature}
version of
     (\ref{e:tadpole}) reads
     \begin{equation}
     I = T\sum_{n=-\infty}^{+\infty} \int \frac{d^3 {\bf
p}}{(2\pi)^3}
     \frac{1}{p^2},
     \label{e:tadpoleT}
     \end{equation}
     with $p_4 = 2\pi nT$.  It is instructive to solve
     Eq.~(\ref{e:tadpoleT})
     by using both the contour method and the zeta-function method.
     The contour method gives \cite{Kapusta,Thesis}:
     \begin{equation}
     I = T\int \frac{d^3 {\bf p}}{(2\pi)^3} \frac{1}{2\pi i}
\oint_c dp_0
     \frac{1}{p_0^{\ 2} - {\bf p}^2 }\frac{1}{2} \beta
     \coth\big (\frac{\beta p_0}{2} \big ), \ \ \beta=1/kT, k=1,
     \label{e:tt}
     \end{equation}
     or
     \begin{eqnarray}
     I &=& T\beta \int \frac{d^3 {\bf p}}{(2\pi)^3} \Bigg \lbrace
     \frac{1}{2\pi i}
     \int_{i\infty-\epsilon}^{-i\infty-\epsilon} dp_0
     \frac{1}{p_0^{\ 2} - {\bf p}^2} \Big (-\frac{1}{2}
     -\frac{1}{e^{-\beta p_0}-1} \Big ) \nonumber \\
     && \ \ \ \ \ + \frac{1}{2\pi i}
     \int_{-i\infty+\epsilon}^{i\infty+\epsilon}
     dp_0 \frac{1}{p_0^{\ 2} - {\bf p}^2}
     \Big (\frac{1}{2} + \frac{1}{e^{\beta p_0}-1} \Big )
     \Bigg \rbrace, \nonumber \\
     &=&I^{{}^{\mbox{vac}}} + I^{{}^{\mbox{matt}}},
     \end{eqnarray}
     where
     \begin{equation}
     I^{{}^{\mbox{vac}}} = \int \frac{d^4 p}{(2\pi)^4} \Big (
     \frac{1}{p_4^{\ 2}
     + {\bf p}^2} \Big ) = 0, \ \ p_0 = ip_4,
     \end{equation}
     and
     \begin{eqnarray}
     I^{{}^{\mbox{matt}}} &=& \int \frac{d^3 {\bf p}}{(2\pi)^3}
     \frac{1}{|{\bf p}|}
     \frac{1}{e^{\beta |{\bf p}|} - 1}, \nonumber \\
     &=& \frac{4\pi}{(2\pi)^3} \int_0^{\infty} dp \frac{p}{e^{\beta
p} -
     1},
     \ \ \ \mbox{ (integrating over the angles)} \nonumber \\
     &=& \frac{4\pi}{(2\pi)^3} \Gamma(2) \zeta(2)
\frac{1}{\beta^2},
     \nonumber \\
     &=& \frac{T^2}{12}.
     \label{e:131}
     \end{eqnarray}
     Thus $I=T^2/12$.

     Next we compute the tadpole in Eq.~(\ref{e:tadpoleT}) by
employing
the method
     of
     zeta-functions \cite{Brandt}.  Making a Wick rotation to
Euclidean
     space and
     using the Schwinger representation for the denominators, we
get
     \begin{eqnarray}
     I&=&T\sum_{n=-\infty}^{+\infty}\int_0^{\infty} d\alpha
e^{-\alpha
     p_4^{\ 2}}
     \int \frac{d^3 {\bf p}}{(2\pi)^3} e^{-\alpha {\bf p}^2},
\nonumber \\
     &=& T\sum_{n=-\infty}^{+\infty}\int_0^{\infty} d\alpha
e^{-\alpha
     p_4^{\ 2}}
     \frac{\pi^{\omega-1/2}}{(2\pi)^{2\omega-1}}
\alpha^{1/2-\omega},
     \nonumber \\
     &=& \frac{1}{8\pi^{3/2}} \Gamma(-\frac{1}{2}) T
     \sum_{n=-\infty}^{+\infty}
     |p_4|, \ \ \  \ \omega = 2. \label{e:rrr}
     \end{eqnarray}
     The sum over $|p_4|$ looks divergent, but is actually finite
     in the spirit of analytic continuation.
     The relevant analytic continuations of the gamma and zeta
functions
     are \cite{Big2}:
     \begin{eqnarray}
     &&\zeta(1-\alpha) = \pi^{-\alpha} 2^{1-\alpha} \Gamma(\alpha)
     \zeta(\alpha)
     \cos\big ( \frac{\pi \alpha}{2} \big ), \label{e:132} \\
     && \Gamma\big (-m+\frac{1}{2} \big )  = \frac{(-1)^m 2^m
\sqrt{\pi}}{
     1\cdot 3 \cdot 5 \cdots (2m-1)}, \ \ \ m=1,2,\ldots.
\label{e:133}
     \end{eqnarray}
     Applying formulas (\ref{e:132}) and (\ref{e:133}) to
     Eq.~(\ref{e:rrr}), we
     see that
     \begin{eqnarray}
     I &=& \frac{1}{8\pi^{3/2}} 4T^2\pi \Gamma \big (-\frac{1}{2}
\big )
     \zeta(-1), \nonumber \\
     &=& \frac{T^2}{12},
     \label{e:134}
     \end{eqnarray}
     since $\Gamma(-1/2) = -2\sqrt{\pi}$ and $\zeta(-1) = -1/12$.
Note
     that
     Eq.~(\ref{e:134}) agrees with Eq.~(\ref{e:131}), and that both
     results vanish in the limit $T \rightarrow 0$.  This answer is
     consistent with the conclusions in refs. \cite{C-L}.

     There are two other integrals which are similar to tadpoles
and arise
     in the
     calculation of the thermodynamic pressure. They are included
here for
     completeness.
     The first of these integrals is calculated as follows:
     \begin{eqnarray}
     T\sum_{n=-\infty}^{+\infty} \int \frac{d^3 {\bf p}}{(2\pi)^3}
     &=& T\sum_{n=-\infty}^{+\infty} \int \frac{d^3 {\bf
p}}{(2\pi)^3}
     \frac{p_4^{\ 2} + {\bf p}^2}{p_4^{\ 2} + {\bf p}^2}, \nonumber
\\
     &=& T\sum_{n=-\infty}^{+\infty} \int_0^{\infty} d\gamma
e^{-\gamma
     p_4^{\ 2}}
     \int \frac{d^3 {\bf p}}{(2\pi)^3} (p_4^{\ 2} + {\bf p}^2)
     e^{-\gamma {\bf p}^2}, \nonumber \\
     &=& 2T \frac{\pi^{3/2}}{(2\pi)^3} \big ( -2\sqrt{\pi} +
2\sqrt{\pi}
     \big )
     (2\pi T)^3 \zeta(-3), \nonumber \\
     &=&  0.
     \end{eqnarray}
     The second integral may be decomposed as follows:
     \begin{equation}
     T\sum_{n=-\infty}^{+\infty} \int \frac{d^3 {\bf p}}{(2\pi)^3}
p^2
     =  I_1 + I_2,
     \end{equation}
     where
     \begin{eqnarray}
     I_1 &\equiv& T\sum_{n=-\infty}^{+\infty} \int \frac{d^3 {\bf
     p}}{(2\pi)^3}
     p_4^{\ 2}, \nonumber \\
     &=& T(2\pi T)^2 2\sum_{n=1}^{\infty} n^2 \int \frac{d^3 {\bf
     p}}{(2\pi)^3},
     \nonumber \\
     &=& T(2\pi T)^2 2\zeta(-2) \int \frac{d^3 {\bf p}}{(2\pi)^3},
     \nonumber \\
     &=& 0, \ \ \zeta(-2) = 0,
     \end{eqnarray}
     while
     \begin{eqnarray}
     I_2 &\equiv& T\sum_{n=-\infty}^{+\infty} \int \frac{d^3
     {\bf p}}{(2\pi)^3} {\bf p}^2,  \nonumber \\
     &=& T\sum_{n=-\infty}^{+\infty} \frac{1}{(2\pi)^3}
\int_0^{\infty}
     d\gamma e^{-\gamma p_4^{\ 2}} \Big ( \frac{-\partial}{\partial
\gamma}
     \Big )
     \int d^3 {\bf p} (p_4^{\ 2} + {\bf p}^2) e^{-\gamma {\bf
p}^2},
     \nonumber \\
     &=& T\sum_{n=-\infty}^{+\infty} \frac{\pi^{3/2}}
     {(2\pi)^3} \int_0^{\infty} d\gamma e^{-\gamma p_4^{\ 2}} \Bigg
(
     p_4^{\ 2} \big ( -3/2 \big ) \gamma^{-5/2}
     + \frac{3}{2} \big ( -5/2 \big )\gamma^{-7/2}
     \Bigg ), \nonumber \\
     &=& T\sum_{n=-\infty}^{+\infty} \frac{1}
     {8\pi^{3/2}} \frac{\Gamma(-3/2)}
     {\big ( p_4^{\ 2} \big )^{-5/2} } \Big [ \frac{1}{2} - 2
     +\frac{3}{2} \Big ], \nonumber \\
     &=& 0.
     \end{eqnarray}
     Thus
     \begin{equation}
     T\sum_{n=-\infty}^{+\infty} \int \frac{d^3 {\bf p}}{(2\pi)^3}
p^2 = 0.
     \end{equation}

     We now have enough machinery to compute all 63 integrals in
     Eq.~(\ref{e:Bigsum}).  Their values are listed in the
Appendix.
     Substituting
     these integrals into Eq.~(\ref{e:Bigsum}), we only need to
complete
     the
     indicated summations in order to derive $P_2^{\mbox{{\small
oyster}}}$.

For
     instance,
     \begin{eqnarray}
     T^2\sum_{n=-\infty}^{+\infty}\sum_{l=-\infty}^{+\infty}
     \Bigg [ \frac{|p_4||k_4|}{16\pi^2} \Bigg ]
     &=& T^2\frac{(2\pi T)^2}{16\pi^2} 2^2 \sum_{n=1}^{\infty}
     \sum_{l=1}^{\infty} nl, \nonumber \\
     &\longrightarrow& T^4 \Big [ \zeta(-1) \Big ]^2 = T^4 \Big [
     \frac{1}{12}
     \Big ]^2.
     \end{eqnarray}
     The final expression for the oyster diagram reads
     \begin{equation}
     P_2^{\mbox{{\small oyster}}} = -7g^2N_cN_g\frac{T^4}{144}.
     \label{e:oyster}
     \end{equation}

     The contribution from the bowtie diagram in Figure 1 is much
easier to
     calculate.  We have
     \begin{eqnarray}
     P_2^{\mbox{{\small bowtie}}} &=& -g^2N_cN_g
     T\sum_{n=-\infty}^{+\infty}
     \int\frac{d^3{\bf p}}{(2\pi)^3} T\sum_{l=-\infty}^{+\infty}
     \int\frac{d^3{\bf k}}{(2\pi)^3} \frac{1}{p^2k^2} \nonumber \\
     &\ &\cdot \big ( 2\delta_{\mu\nu}\delta_{\lambda\rho}
     -\delta_{\mu\lambda}\delta_{\nu\rho} -
     \delta_{\mu\rho}\delta_{\lambda\nu}
     \big ) \nonumber \\
     &\ &\cdot \Bigg ( \delta^{\lambda\rho} +
\frac{p^{\lambda}p^{\rho}}
     {p_4^{\ 2}} \Bigg ) \Bigg ( \delta^{\mu\nu} +
\frac{k^{\mu}k^{\nu}}
     {k_4^{\ 2}} \Bigg ), \ \ \lambda, \rho, \mu, \nu = 1,2,3,
\nonumber \\
     &=& -g^2N_cN_g T\sum_{n=-\infty}^{+\infty}
     \int\frac{d^3{\bf p}}{(2\pi)^3} T\sum_{l=-\infty}^{+\infty}
     \int\frac{d^3{\bf k}}{(2\pi)^3} \frac{1}{p^2k^2} \nonumber \\
     &\ & \cdot \Bigg ( 6 + 2\frac{k^2}{k_4^{\ 2}} +
2\frac{p^2}{p_4^{\ 2}}
     + 2\frac{p^2k^2}{p_4^{\ 2}k_4^{\ 2}} - 2\frac{({\bf p} \cdot
{\bf
     k})^2}
     {p_4^{\ 2}k_4^{\ 2}} \Bigg ), \nonumber \\
     &=& -g^2N_cN_g T^2\sum_{n=-\infty}^{+\infty}
     \sum_{l=-\infty}^{+\infty}
     \frac{10}{3} \Bigg [ \frac{|p_4||k_4|}{16\pi^2} \Bigg ],
\nonumber \\
     &=& -\frac{10}{3}g^2N_cN_g\frac{T^4}{144}.
     \label{e:bowtie}
     \end{eqnarray}
     Combining Eqs.~(\ref{e:oyster}) and (\ref{e:bowtie}) subject
to their
     respective symmetry factors, we find that the pure-gauge
contribution
     to the
     two-loop pressure is given by
     \begin{eqnarray}
     P_2^{\mbox{{\small glue}}} &=& \frac{1}{12} P_2^{\mbox{{\small
     oyster}}}
     + \frac{1}{8} P_2^{\mbox{{\small bowtie}}}, \nonumber \\
     &=& -\frac{g^2}{144} N_cN_g T^4.
     \label{e:last}
     \end{eqnarray}
     This value agrees with the Feynman-gauge result in
     Eq.~(\ref{e:P2Feynman}).

     \section{Conclusion}

     In this article we have used the temporal gauge, $A_0^a(x)=0$,
to
     derive the
     thermodynamic pressure in Q.C.D. at one and two loops.  Our
answer for
     the
     one-loop pressure, Eq.~(\ref{e:catsmeow}), agrees with the
well-known
     formula
     for the black-body radiation, while our two-loop result in
     Eq.~(\ref{e:last})
     turns out to be identical to the Feynman-gauge answer.

     These results were achieved by replacing the traditional
contour
     method by the
     method of zeta-functions, and by exploiting a special version
of the
     unified-gauge prescription to handle the spurious poles of the
gluon
     propagator,
     Eq.~(\ref{e:propagator}).  These spurious poles may be treated
by the
     causal
     prescription \cite{L2} (in Minkowski space):
     \begin{eqnarray}
     \frac{1}{p_0} \vline^{\mbox{temp}} && = \lim_{\epsilon \to 0}
     \frac{p_0}{p_0^2+i\epsilon}, \ \ \epsilon > 0, \nonumber \\
     \frac{1}{(p_0)^2} \vline^{\mbox{temp}} && = \lim_{\epsilon \to
0}
     \Big (\frac{p_0}{p_0^2+i\epsilon}\Big )^2, \ \ \epsilon > 0,
     \end{eqnarray}
     which is well defined for $p_0=2\pi inT=0$, namely at $n=0$.
     Consistent
     application of this prescription enabled us to evaluate
unambiguously
     all
     63 two-loop integrals in the Appendix.

     \vspace{7 mm}
     \begin{center}
     {\bf {\Large Acknowledgments}}
     \end{center}
     It is a pleasure to thank Gabor Kunstatter for several
discussions and
     for
     referring us to reference [11].  One of us (M.S.) would like
to
     acknowledge
     financial support from the Natural Sciences and Engineering
Research
     Council
     of Canada.  This research was supported in part by the Natural
     Sciences and
     Engineering Research Council of Canada under grant No. A8063.

     \newpage
     \begin{center}
     {\bf {\Large Appendix}}
     \end{center}

     We list here the 63 integrals \cite{Thesis} needed in the
computation
     of
     $P_2^{\mbox{oyster}}$ in Eq.~(\ref{e:Bigsum}).
     \begin{eqnarray}
     &&I_1 = \int\int{\frac {1}{{  p_4}\,\left ({  p_4}+{
k_4}\right
     )^{2}{  k_4}}}
     = 0, \ \ \ \ \
     \int\int \equiv \int\frac{d^3{\bf p}}{(2\pi)^3}
     \int\frac{d^3{\bf k}}{(2\pi)^3}, \nonumber \\
     &&I_2 = -\int\int{\frac {{  p_4}}{\left ({  p_4}+{  k_4}\right
)^{2}{
     k_4}\,{
     (k+p)^2}}}
     = 0, \nonumber \\
     &&I_3 = +\int\int{\frac {21\,{  k_4}^{2}}{\left ({  p_4}+{
k_4}\right
     )^{2}{
       k^2}\,{  (k+p)^2}}}
     = 21\frac{k_4^{\ 2}}{p_4^{\ 2}} \Bigg [
\frac{|p_4||k_4|}{16\pi^2}
     \Bigg ],
     \nonumber \\
     &&I_4 = +\int\int{\frac {33\,{  p_4}^{2}}{\left ({  p_4}+{
k_4
     }\right )^{2}{  k^2}\,{  (k+p)^2}}}=
33\frac{(p_4-k_4)^2}{p_4^{\ 2}}
     \Bigg [ \frac{|p_4||k_4|}{16\pi^2} \Bigg ], \nonumber \\
     &&I_5 = +\int\int{\frac {42\,{  p_4}\,{  k_4}}{
     \left ({  p_4}+{  k_4}\right )^{2}{  k^2}\,{  (k+p)^2}}}
     = 42 \frac{k_4(p_4-k_4)}{p_4^{\ 2}}
     \Bigg [ \frac{|p_4||k_4|}{16\pi^2} \Bigg ],  \nonumber \\
     &&I_6 = +\int\int{\frac {5\,{
       k^2}}{\left ({  p_4}+{  k_4}\right )^{2}{  p^2}\,{
(k+p)^2}}}
     = -10\frac{p_4}{k_4} \Bigg [ \frac{|p_4||k_4|}{16\pi^2} \Bigg
] 0,
     \nonumber \\
     &&I_7 = -\int\int{\frac {{  p_4}^{2}}{\left ({  p_4}+{
k_4}\right
     )^{2}{  k_4}^{2}{
       (k+p)^2}}}= 0, \nonumber \\
     &&I_8 = -\int\int{\frac {27\,{  p_4}^{2}}{\left ({  p_4}+{
k_4}\right
     )^
     {2}{  p^2}\,{  k^2}}}= -27\frac{p_4^{\ 2}}{(p_4+k_4)^2}
     \Bigg [ \frac{|p_4||k_4|}{16\pi^2} \Bigg ], \nonumber \\
     &&I_9 = -\int\int{\frac {9}{\left ({  p_4}+{  k_4}\right )^{2
     }{  k^2}}}= 0, \nonumber \\
     &&I_{10} = -\int\int{\frac {12\,{  k_4}}{{  p_4}\,\left ({
p_4}+{
     k_4}
     \right )^{2}{  p^2}}}= 0, \nonumber \\
     &&I_{11} = +\int\int{\frac {{  (k+p)^2}}{{  p_4}\,\left ({
p_4}+{
       k_4}\right )^{2}{  k_4}\,{  k^2}}}
     = 0, \nonumber \\
     &&I_{12} = -\int\int{\frac {{  p^2}}{2\,{  p_4}
     \,\left ({  p_4}+{  k_4}\right )^{2}{  k_4}\,{  k^2}}}= 0,
\nonumber
     \\
     &&I_{13} = +\int\int{\frac {{
       (k+p)^2}}{4\,\left ({  p_4}+{  k_4}\right )^{2}{  k_4}^{2}{
k^2}}}
     = 0, \nonumber \\
     &&I_{14} = +\int\int{\frac {{  (k+p)^2}}{2\,{  p_4}^{2}\left
({
     p_4}+{  k_4}\right )^{
     2}{  k^2}}}= 0, \nonumber \\
     &&I_{15} = -\int\int{\frac {{  k^2}}{4\,\left ({  p_4}+{
k_4}\right
     )^{2}
     {  k_4}^{2}{  (k+p)^2}}}= 0, \nonumber \\
     &&I_{16} = -\int\int{\frac {{  k^2}}{2\,{  p_4}\,\left ({
p_4}
     +{  k_4}\right )^{2}{  k_4}\,{  p^2}}}= 0, \nonumber \\
     &&I_{17} = +\int\int{\frac {42\,{  p_4}\,{
     k_4}}{\left ({  p_4}+{  k_4}\right )^{2}{  p^2}\,{  (k+p)^2}}}
     = 42\frac{p_4(k_4-p_4)}{k_4^{\ 2}}
     \Bigg [ \frac{|p_4||k_4|}{16\pi^2} \Bigg ],  \nonumber \\
     &&I_{18} = +\int\int{\frac
     {1}{\left ({  p_4}+{  k_4}\right )^{2}{  k_4}^{2}}}
     = 0, \nonumber \\
     &&I_{19} = +\int\int{\frac {1}{{
       p_4}^{2}\left ({  p_4}+{  k_4}\right )^{2}}}= 0, \nonumber
\\
     &&I_{20} = -\int\int{\frac {6}{\left ({
       p_4}+{  k_4}\right )^{2}{  (k+p)^2}}}= 0, \nonumber \\
     &&I_{21} = -\int\int{\frac {30\,{  p_4}\,{  k_4}
     }{\left ({  p_4}+{  k_4}\right )^{2}{  p^2}\,{  k^2}}}
     = -30\frac{p_4k_4}{(p_4+k_4)^2}
     \Bigg [ \frac{|p_4||k_4|}{16\pi^2} \Bigg ], \nonumber \\
     &&I_{22} = -\int\int{\frac {27
     \,{  k_4}^{2}}{\left ({  p_4}+{  k_4}\right )^{2}{  p^2}\,{
k^2}
     }}
     = -27\frac{k_4^{\ 2}}{(p_4+k_4)^2}
     \Bigg [ \frac{|p_4||k_4|}{16\pi^2} \Bigg ], \nonumber \\
     &&I_{23} = -\int\int{\frac {3\,{  k_4}}{{  p_4}\,\left ({
p_4}+{
     k_4}\right )^{2}
     {  k^2}}}= 0, \nonumber \\
     &&I_{24} = +\int\int{\frac {8\,{  (k+p)^2}}{\left ({  p_4}+{
     k_4}\right )^{2}{
       p^2}\,{  k^2}}}= 16\frac{p_4k_4}{(p_4+k_4)^2}
       \Bigg [ \frac{|p_4||k_4|}{16\pi^2} \Bigg ], \nonumber \\
     &&I_{25} = +\int\int{\frac {3\,{  p_4}^{2}{  (k+p)^2}}{\left
({  p_4}+
     {  k_4}\right )^{2}{  k_4}^{2}{  p^2}\,{  k^2}}}
     = 6\frac{p_4^{\ 3}k_4}{(p_4+k_4)^2k_4^{\ 2}}
     \Bigg [ \frac{|p_4||k_4|}{16\pi^2} \Bigg ], \nonumber \\
     &&I_{26} = +\int\int{\frac {5\,{
     p_4}\,{  p^2}}{\left ({  p_4}+{  k_4}\right )^{2}{  k_4}\,{
k^2}
     \,{  (k+p)^2}}}
     = -10\frac{(p_4-k_4)p_4}{p_4^{\ 2}}
     \Bigg [ \frac{|p_4||k_4|}{16\pi^2} \Bigg ], \nonumber \\
     &&I_{27} = +\int\int{\frac {12\,{  p_4}^{3}}{\left ({  p_4}+{
k_4}
     \right )^{2}{  k_4}\,{  k^2}\,{  (k+p)^2}}}
     = 12\frac{(p_4-k_4)^3}{p_4^{\ 2}k_4}
     \Bigg [ \frac{|p_4||k_4|}{16\pi^2} \Bigg ], \nonumber \\
     &&I_{28} = +\int\int{\frac {7\,{  p_4}\,{
     (k+p)^2}}{\left ({  p_4}+{  k_4}\right )^{2}{  k_4}\,{
p^2}\,{  k^2}
     }}= 14\frac{p_4^{\ 2}}{(p_4+k_4)^2}
     \Bigg [ \frac{|p_4||k_4|}{16\pi^2} \Bigg ], \nonumber \\
     &&I_{29} = -\int\int{\frac {12\,{  p_4}^{3}}{\left ({  p_4}+{
     k_4}\right )^{2}{
     k_4}\,{  p^2}\,{  k^2}}}
     = -12\frac{p_4^{\ 2}}{(p_4+k_4)^2k_4}
     \Bigg [ \frac{|p_4||k_4|}{16\pi^2} \Bigg ], \nonumber \\
     &&I_{30} = -\int\int{\frac {6\,{  p_4}^{2}}{\left ({  p_4}+{
       k_4}\right )^{2}{  k_4}^{2}{  k^2}}}= 0, \nonumber \\
     &&I_{31} = -\int\int{\frac {{  k^2}}{4\,\left (
     {  p_4}+{  k_4}\right )^{2}{  k_4}^{2}{  p^2}}}= 0, \nonumber
\\
     &&I_{32} = -\int\int{\frac {{  (k+p)^2}^
     {2}}{4\,\left ({  p_4}+{  k_4}\right )^{2}{  k_4}^{2}{
p^2}\,{
       k^2}}}= -\frac{4}{3}\frac{p_4^{\ 2}}{(p_4+k_4)^2}
       \Bigg [ \frac{|p_4||k_4|}{16\pi^2} \Bigg ], \nonumber \\
     &&I_{33} = -\int\int{\frac {{  p^2}^{2}}{4\,\left ({  p_4}+{
     k_4}\right )^{2
     }{  k_4}^{2}{  k^2}\,{  (k+p)^2}}}
     = -\frac{4}{3}\Bigg [ \frac{|p_4||k_4|}{16\pi^2} \Bigg ],
\nonumber
     \\
     &&I_{34} = +\int\int{\frac {{  k^2}}{4\,{  p_4}^{2}
     \left ({  p_4}+{  k_4}\right )^{2}{  p^2}}}= 0, \nonumber \\
     &&I_{35} = +\int\int{\frac {{  p^2}}{4\,
     \left ({  p_4}+{  k_4}\right )^{2}{  k_4}^{2}{  k^2}}}= 0,
\nonumber
     \\
     &&I_{36} = -\int\int{\frac {{
       p^2}}{4\,{  p_4}^{2}\left ({  p_4}+{  k_4}\right )^{2}{
k^2}}}
     = 0, \nonumber \\
     &&I_{37} = +\int\int{\frac {{  (k+p)^2}}{2\,\left ({  p_4}+{
     k_4}\right )^{2}{  k_4}^{2
     }{  p^2}}}= 0, \nonumber \\
     &&I_{38} = +\int\int{\frac {{  (k+p)^2}}{4\,{  p_4}^{2}\left
({
     p_4}+{  k_4}
     \right )^{2}{  p^2}}}= 0, \nonumber \\
     &&I_{39} = -\int\int{\frac {6\,{  k_4}^{2}}{{  p_4}^{2}\left
({
       p_4}+{  k_4}\right )^{2}{  p^2}}}= 0, \nonumber \\
     &&I_{40} = +\int\int{\frac {12\,{  k_4}^{3}}{{
     p_4}\,\left ({  p_4}+{  k_4}\right )^{2}{  p^2}\,{  (k+p)^2}}}
     = 12\frac{(k_4-p_4)^3}{p_4k_4^{\ 2}}
     \Bigg [ \frac{|p_4||k_4|}{16\pi^2} \Bigg ], \nonumber \\
     &&I_{41} = +\int\int{\frac
     {5\,{  k_4}\,{  k^2}}{{  p_4}\,\left ({  p_4}+{  k_4}\right
)^{2}
     {  p^2}\,{  (k+p)^2}}}= -10\frac{(k_4-p_4)}{k_4}
     \Bigg [ \frac{|p_4||k_4|}{16\pi^2} \Bigg ], \nonumber \\
     &&I_{42} = -\int\int{\frac {9}{\left ({  p_4}+{  k_4}\right
)^{2}{
       p^2}}}= 0, \nonumber \\
     &&I_{43} = +\int\int{\frac {{  k^2}}{2\,{  p_4}^{2}\left ({
p_4}+{
     k_4}
     \right )^{2}{  (k+p)^2}}}= 0, \nonumber \\
     &&I_{44} = +\int\int{\frac {{  p^2}}{2\,\left ({  p_4}+{  k_4}
     \right )^{2}{  k_4}^{2}{  (k+p)^2}}}= 0, \nonumber \\
     &&I_{45} = -\int\int{\frac {{  p^2}}{4\,{  p_4}^{2}
     \left ({  p_4}+{  k_4}\right )^{2}{  (k+p)^2}}}= 0, \nonumber
\\
     &&I_{46} = +\int\int{\frac {5\,{  p^2}}{
     \left ({  p_4}+{  k_4}\right )^{2}{  k^2}\,{  (k+p)^2}}}
     = -10\frac{k_4}{p_4}\Bigg [ \frac{|p_4||k_4|}{16\pi^2} \Bigg
],
     \nonumber \\
     &&I_{47} = +\int\int{\frac {21\,
     {  p_4}^{2}}{\left ({  p_4}+{  k_4}\right )^{2}{  p^2}\,{
(k+p)^2}}}
     = 21\frac{p_4^{\ 2}}{k_4^{\ 2}}
     \Bigg [ \frac{|p_4||k_4|}{16\pi^2} \Bigg ], \nonumber \\
     &&I_{48} = +\int\int{\frac {33\,{  k_4}^{2}}{\left ({  p_4}+{
     k_4}\right )^{2}{
     p^2}\,{  (k+p)^2}}}= 33\frac{(k_4-p_4)^2}{k_4^{\ 2}}
     \Bigg [ \frac{|p_4||k_4|}{16\pi^2} \Bigg ], \nonumber \\
     &&I_{49} = -\int\int{\frac {12\,{  k_4}^{3}}{{  p_4}\,\left ({
p_4}+{
       k_4}\right )^{2}{  p^2}\,{  k^2}}}
     = -12\frac{k_4^{\ 3}}{p_4(p_4+k_4)^2}
     \Bigg [ \frac{|p_4||k_4|}{16\pi^2} \Bigg ], \nonumber \\
     &&I_{50} = +\int\int{\frac {7\,{  k_4}\,{  (k+p)^2}
     }{{  p_4}\,\left ({  p_4}+{  k_4}\right )^{2}{  p^2}\,{
k^2}}}
     = 14\frac{k_4^{\ 2}}{(p_4+k_4)^2}
     \Bigg [ \frac{|p_4||k_4|}{16\pi^2} \Bigg ], \nonumber \\
     &&I_{51} = -\int\int{\frac {{  k^2}^{2}}{4\,{  p_4}^{2}\left
({
     p_4}+{  k_4}\right )^{
     2}{  p^2}\,{  (k+p)^2}}}
     = -\frac{4}{3}\Bigg [ \frac{|p_4||k_4|}{16\pi^2} \Bigg ],
\nonumber
     \\
     &&I_{52} = -\int\int{\frac {3\,{  p_4}}{\left ({  p_4}+{  k_4}
     \right )^{2}{  k_4}\,{  p^2}}}= 0, \nonumber \\
     &&I_{53} = -\int\int{\frac {{  k_4}^{2}}{{  p_4}^{2}
     \left ({  p_4}+{  k_4}\right )^{2}{  (k+p)^2}}}= 0, \nonumber
\\
     &&I_{54} = +\int\int{\frac {{  (k+p)^2}}{{
       p_4}\,\left ({  p_4}+{  k_4}\right )^{2}{  k_4}\,{  p^2}}}
     = 0, \nonumber \\
     &&I_{55} = -\int\int{\frac {{  (k+p)^2}^{2}}{2\,{  p_4}\,\left
({
     p_4}+{  k_4}\right )^{2
     }{  k_4}\,{  p^2}\,{  k^2}}}=
-\frac{8}{3}\frac{p_4k_4}{(p_4+k_4)^2}
     \Bigg [ \frac{|p_4||k_4|}{16\pi^2} \Bigg ], \nonumber \\
     &&I_{56} = +\int\int{\frac {3\,{  p_4}^{2}{  p^2}}{
     \left ({  p_4}+{  k_4}\right )^{2}{  k_4}^{2}{  k^2}\,{
(k+p)^2}}}
     = -6\frac{(p_4-k_4)^2}{p_4k_4}
     \Bigg [ \frac{|p_4||k_4|}{16\pi^2} \Bigg ], \nonumber \\
     &&I_{57} = \int\int{\frac {3\,{  k_4}^{2}{  k^2}}{{
p_4}^{2}\left ({
     p_4}+{  k_4}
     \right )^{2}{  p^2}\,{  (k+p)^2}}}
     =  -6\frac{(k_4-p_4)^2}{p_4k_4}
     \Bigg [ \frac{|p_4||k_4|}{16\pi^2} \Bigg ], \nonumber \\
     &&I_{58} = +\int\int{\frac {3\,{  k_4}^{2}{  (k+p)^2}}{{
     p_4}^{2}\left ({  p_4}+{  k_4}\right )^{2}{  p^2}\,{  k^2}}}
     = 6\frac{k_4^{\ 3}}{p_4(p_4+k_4)^2}
     \Bigg [ \frac{|p_4||k_4|}{16\pi^2} \Bigg ], \nonumber \\
     &&I_{59} = -\int\int{\frac {{  k_4}^{2}}{{  p_4}^{2}\left ({
p_4}+{
     k_4}\right )^{2}{
       k^2}}}= 0, \nonumber \\
     &&I_{60} = -\int\int{\frac {{  k_4}}{{  p_4}\,\left ({  p_4}+{
     k_4}\right )
     ^{2}{  (k+p)^2}}}= 0, \nonumber \\
     &&I_{61} = -\int\int{\frac {{  (k+p)^2}^{2}}{4\,{
p_4}^{2}\left ({
     p_4}+
     {  k_4}\right )^{2}{  p^2}\,{  k^2}}}
     = -\frac{4}{3}\frac{k_4^{\ 2}}{(p_4+k_4)^2}
     \Bigg [ \frac{|p_4||k_4|}{16\pi^2} \Bigg ], \nonumber \\
     &&I_{62} = -\int\int{\frac {12\,{  p_4}}{\left (
     {  p_4}+{  k_4}\right )^{2}{  k_4}\,{  k^2}}}= 0, \nonumber \\
     &&I_{63} = -\int\int{\frac {{  p_4}^{2}
     }{\left ({  p_4}+{  k_4}\right )^{2}{  k_4}^{2}{  p^2}}} =0.
\nonumber
     \end{eqnarray}

     \vfill\eject

     \newpage
     \center{{\bf {\Large Figure Caption}}}

     Figure 1. {\em Diagrams contributing to the thermodynamic
pressure in
     Q.C.D}


\begin{thebibliography}{99}

     \bibitem{Weyl}H. Weyl, Z. Phys. {\bf 56} (1929) 330.
     \bibitem{Heisenberg}W. Heisenberg and W. Pauli, {\em Z. Phys.}
{\bf
     59} (1930) 168.
     \bibitem{Rossi}G.C. Rossi and M. Testa, {\em Phys. Rev.} {\bf
D29}
     (1984) 2997.
     \bibitem{Claudson}M. Claudson and M.B. Halpern, {\em Phys.
Lett.} {\bf
     B151} (1985) 281.
     \bibitem{Bern}Z. Bern and M.S. Chan, {\em Nucl. Phys.} {\bf
B266}
     (1986) 509.
     \bibitem{Actor}A. Actor, {\em in} \underbar{New perspectives
in
     quantum field theories},
     eds. J. Abad, M. Asorey and A. Cruz (World Scientific,
Singapore,
     1986) p. 345.
     \bibitem{Kajantie}K. Kajantie and J.I. Kapusta, {\em Ann.
Phys.}
     (N.Y.)
     {\bf 160} (1985) 477.
     \bibitem{James}K.A. James and P.V. Landshoff, {\em Phys.
Lett.} {\bf
     B251} (1990) 167.
     \bibitem{Brandt}F.T. Brandt, J. Frenkel and J.C. Taylor, {\em
Phys.
     Rev.} {\bf D44}
     (1991) 1801.
     \bibitem{L-S}G. Leibbrandt and M. Staley, {\em Nucl. Phys.}
{\bf B428}
     (1994) 469.
     \bibitem{Rebhan}A.K. Rebhan, {\em in} \underbar{Proceedings
of the 3rd Workshop on Thermal Field Theories and} \underbar{Their
Applications}, eds. F.C.Khanna, R. Kobes, G. Kunstatter and H.
Umezawa (World Scientific, Singapore, 1994) p. 469.
     \bibitem{L1}G. Leibbrandt, {\em Rev. Mod. Phys.} {\bf 59}, No.
4
     (1987) 1067.
     \bibitem{L2}G. Leibbrandt, \underbar{Noncovariant Gauges}
(World
     Scientific,
     Singapore, 1994) p. 62.
     \bibitem{L3}G. Leibbrandt, {\em Nucl. Phys.} {\bf B310} (1988)
405.
     \bibitem{Kapusta}J.I. Kapusta, \underbar{Finite Temperature
Field
     Theory}
     (Cambridge University Press, Cambridge, 1989).
     \bibitem{Big1} I.S. Gradshteyn and I.M. Ryzhik,
\underbar{Table of
     Integrals, Series
     and Products} (Academic Press Inc., San Diego, 1980).
     \bibitem{Kapusta79} J.I. Kapusta, {\em Nucl. Phys.}
     {\bf B148}, 461 (1979).
     \bibitem{C-L}D.M. Capper and G. Leibbrandt, {\em Lettere Nuovo
     Cimento}
     {\bf 6}, 117 (1973);  D.M. Capper and G. Leibbrandt, {\em J.
Math.
     Phys.}
     {\bf 15}, 82 (1974);  {\em J.
Math.
     Phys.}
     {\bf 15}, 86 (1974); {\em J.
Math.
     Phys.}
     {\bf 15}, 795 (1974).
     \bibitem{Thesis}M. D. Staley, \underbar{Finite-Temperature
Q.C.D. in the Temporal Gauge}, Ph.D. Thesis, University of Guelph
(Guelph, 1995, unpublished).
     \bibitem{Big2} M. Abramowitz and I.A. Stegun,
\underbar{Handbook of
     Mathematical
     Functions} (National Bureau of Standards, Washington, 1972).
     \end{thebibliography}
     \end{document}